# A liquid xenon TPC for a medical imaging Compton telescope

**T. Oger, W-T. Chen, J-P. Cussonneau, J. Donnard, S. Duval, J. Lamblin, O. Lemaire, A.F. Mohamad Hadi, P. Leray, E. Morteau, L. Scotto Lavina, J-S. Stutzmann and D. Thers.**

*SUBATECH, UMR 6457, Ecole des Mines de Nantes, CNRS/IN2P3 and Université de Nantes, 44307 Nantes, France*

Abstract

A new technique for medical imaging, "3γ imaging", is studied by our group at SUBATECH for few years. A small liquid xenon time projection chamber prototype has been built in order to demonstrate the feasibility of this technique. With an ultra-low-noise front-end electronics, the energy deposit and resolution of 511 keV γ-ray as a function of drift electric field (E) is measured with high precision. 500 μm of z resolution is estimated by measuring the charge carriers drift velocity and time resolution.

## 1. Introduction

Positron emission tomography (PET) is a nuclear medical imaging technique which is popularly used in oncology; it allows tracing the radioactive sources fixed in a targeted tissue. In a standard PET device, the location of $\beta^+$ emitters is based on the detection in coincidence of the two back-to-back annihilation γ-rays. Hence, the positions of the radioactive sources can only be located on the intersections of a huge amount of line-of-response (LOR). To precisely locate the radioactive sources in real time and in three dimensions, SUBATECH has proposed a novel medical imaging technique, called 3γ imaging. It uses a specific radioisotope, $^{44}$Sc, which emits a γ ($E_\gamma$ =1.157 MeV) and a positron in quasi-coincidence [1]. The principle of the 3γ imaging technique is illustrated in the Figure 1. Two back-to-back 511 keV γ-rays resulting from the positron annihilation can be detected by a classical PET device and create a LOR. The additional γ-ray can be detected by a Compton telescope. The energy deposit of the first hit ($E_1$) in the Compton telescope leads to the determination of the scattering angle θ (1),

$$\cos\theta = 1 - m_e c^2 \frac{E_1}{E_0(E_0 - E_1)} \qquad (1)$$

where $E_0$ is the energy of incident γ-ray (1.157MeV) and $m_e c^2$ is the energy of electron mass (0.511 MeV). The vector reconstructed from the positions of the first and the second hit indicates the axis (Δ) of the Compton cone, which has an aperture angle θ. Therefore, the position of the radioactive source can then be located by calculating the intersection of the Compton cone and the LOR. This concept permits a real-time detection, reaching high image resolutions with much less radiation dose released to the patient.

In order to achieve high performances with the 3γ imaging technique, the Compton telescope must collect Compton scattering events with high efficiency, have a good spatial and a good energy resolution on each interaction vertex of the γ-ray detection. The excellent properties of liquid xenon (LXe) for γ-ray detection make it an attractive radiation detector medium [2]. Its high atomic number (54), high density (~3 g/cm$^3$) and homogeneous volume make it very efficient for stopping penetrating radiations. The mass attenuation coefficient for 1.157 MeV energy γ-ray in LXe is about 0.05 $cm^2/g$, corresponding to ~7 cm of mean-free-path. With a thickness of 12 cm, the probability for a 1.157 MeV γ-rays to undergo at least one Compton scattering in LXe is 79%. As a scintillating material, LXe not only provides high stopping power benefit in a single large and homogeneous volume, but also gives a fast decay time (2.2 ns) of the scintillation signal, which can provide a precise time determination. Concerning the detection of radiation by measuring the ionization signal, LXe has a quite small W-value (15.6 eV [3], defined as the average energy required to produce an electron-ion pair) providing a large ionization yield, and a small Fano Factor (0.041 [4]) leading to potentially relatively small fluctuation of the number of electron-ion pair generated by energy deposit. However, ionizing electrons rapidly recombine to xenon atoms immediately and convert the energy to heat or scintillation, unless they escape because of an external force to separate them. Ionizing electrons are collected on the anode thanks to an electric field applied through the detector, and escape to recombination.

In order to understand the performance of the 3γ imaging using a LXe Compton telescope, we first performed GEANT4 [5] simulations with three major hypotheses concerning the Compton telescope: a 500 µm spatial resolution, an energy resolution $\sigma = \frac{2.3\%}{\sqrt{E_M}}$ where $E_M$ is the energy deposit in MeV, and a front-end electronics with 200 electrons (~10 keV) noise level. The result of simulations shows that a spatial resolution of ~1 cm can be achieved with 3γ imaging for each event without any iterative algorithmic reconstruction.

In the present work, the measurement of charge yield and its resolution as a function of applied electric field is performed. The corresponding drift velocity of ionizing electrons is also measured allowing the estimation of the z (along the drift direction) resolution by multiplying drift velocity and time resolution.

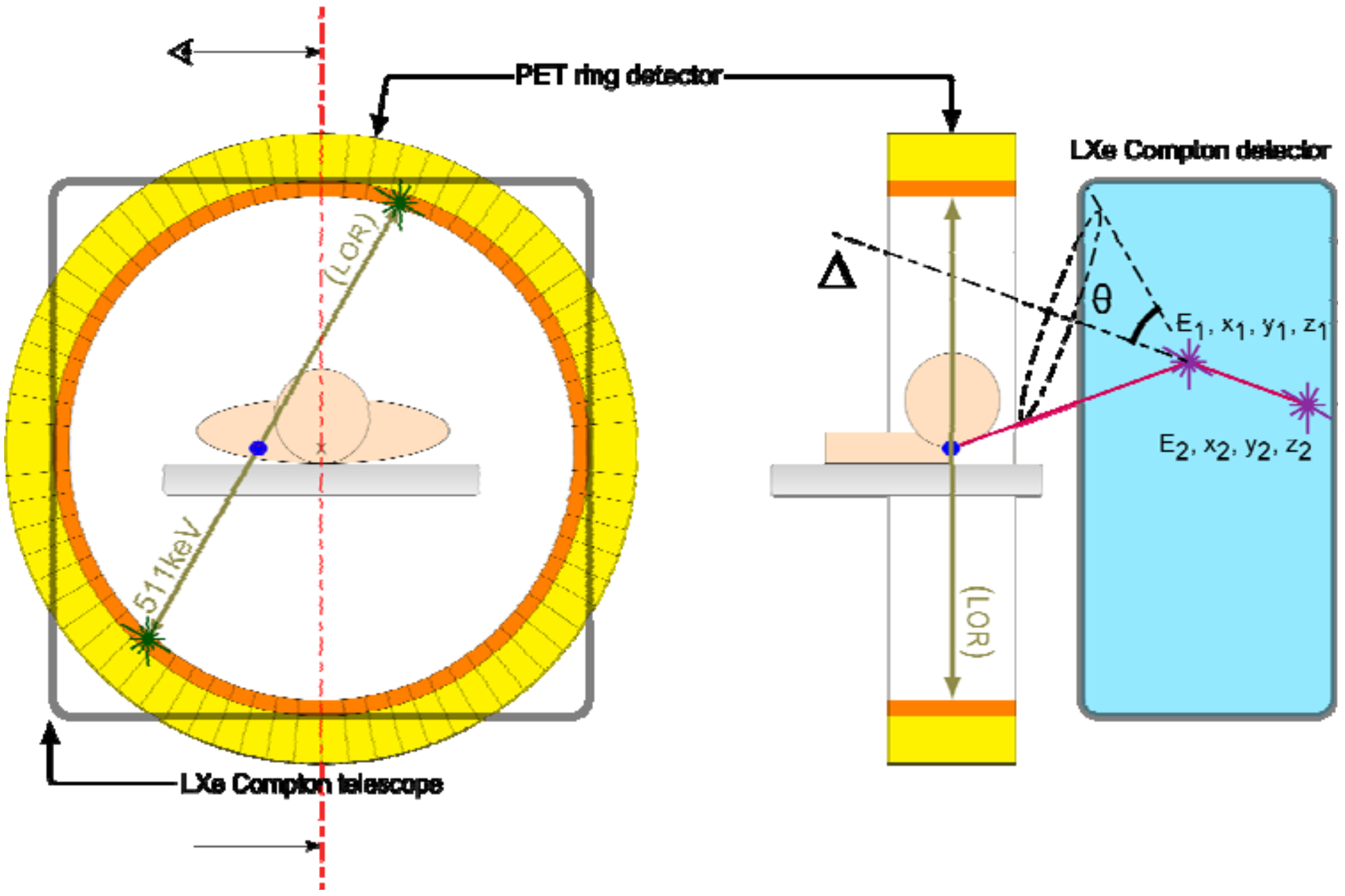

**Figure 1 : Schematic drawing of the 3γ imaging principle**

## 2. Setup and procedure

A small prototype of LXe Time Projection Chamber (TPC), named XEMIS (Xenon Medical Imaging System), has been built in order to validate the 3γ imaging concept by reaching desired technical performance. Due to its small active volume, an effective Compton event tracking cannot be done. The whole apparatus is presented in Figure 2. The LXe TPC is enclosed in a double-walled insulated stainless steel cryostat with heat loss around 46 W when the cryostat is cold. A pulse tube refrigerator (PTR) – Iwatani PC150 – optimized for the use of xenon cryogenics is used to liquefy the xenon and maintain it at around 171 K under a pressure of 1.2 bar. The distance between the PTR and the cryostat is around two meters in order to reduce mechanical vibrations. A purification circuit is used to continuously purify the xenon since the purity of LXe is required to be as high as possible to prevent charge carriers capture by electronegative impurities ($H_2O$, $O_2$) released by construction materials into the liquid. The purification circuit is composed of an Enomoto MX-808ST-S micro-pump to drive the recirculation, a SAES PS4-MT3-R/N getter to purify the xenon, a flow controller to stabilize the flow rate, and a plate heat exchanger, which have to recuperate some of the vapor heat of LXe to liquefy the gaseous xenon (GXe). With heat loss around 46 W and 100 W cooling power produced by PTR, the flow rate of the recirculation can reach ~4.5 L/min.

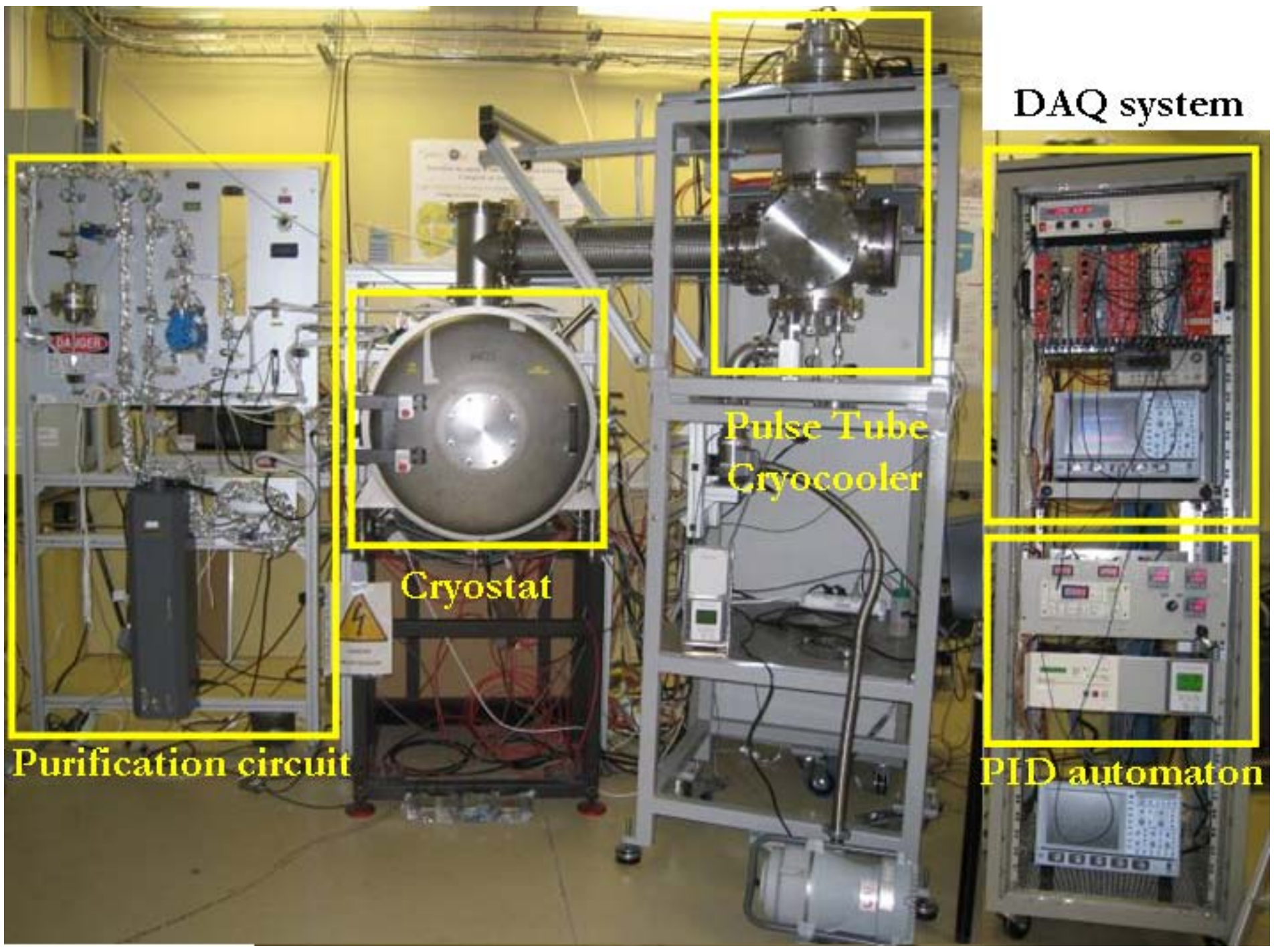

**Figure 2 : XEMIS cryogenic apparatus**

The structure of TPC is presented on the Figure 3. A Hamamatsu R7600-06MOD-ASSY photomultiplier tube (PMT) is used to detect VUV scintillation photons (178 nm). A high optical transparency (90%, 70 lpi) grid is placed in front of the PMT with 1 mm distance used as the cathode. Twelve copper field shaping rings are installed around the TPC in order to provide a uniform electric field within the cylindrical active volume (diameter = 3.6 cm and length = 12 cm), which can be adjusted from 0.5 to 1.2 kV/cm$^2$. To read charge signal, a Micromegas device is used. It consists in a low-optical transparency micromesh (670 lpi), which is used as a Frisch grid, placed at a distance of 300 μm from the anode. To ensure a 100% micromesh transparency of charge carriers, the strength of electric field between the micromesh and anode has to be at least 50 times higher than the drift field applied between the micromesh and the cathode. The anode has an active surface of 2.54x2.54 cm$^2$ and is segmented in 16 pixels. An ultra-low-noise front-end electronics (FEE) with 16 channels, IDeF-X v1.0 ASIC [6], is used for the charge readout. The electronics noise is reduced to ~100 electrons because the FEE is placed just on the other side of the anode and work in the vacuum at LXe temperature (171 K).

In order to characterize the performance of the LXe TPC, a $^{22}$Na source, which is a $\beta^+$-$\gamma$ emitter ($E_{max}$ $\beta^+$= 545 keV, $E_\gamma$ = 1.257 MeV), is placed at 12.5 cm far from the anode. A CsI crystal coupled to a PMT is used to trigger the events which have two back-to-back 511 keV $\gamma$-rays due to positron annihilation. A collimator placed between $^{22}$Na source and CsI crystal can guarantee that one of the back-to-back $\gamma$-rays enters the active volume. The waveforms of PMT signal (scintillation) and anode signals (ionization) are stored when both PMTs deliver a signal. PMT signal is recorded with a CAEN V1720 acquisition card with a sampling frequency of 250 MHz during 4.096 µs. Anode signals are recorded with a CAEN V1740 card at 12.5 MHz during 122.88 µs.

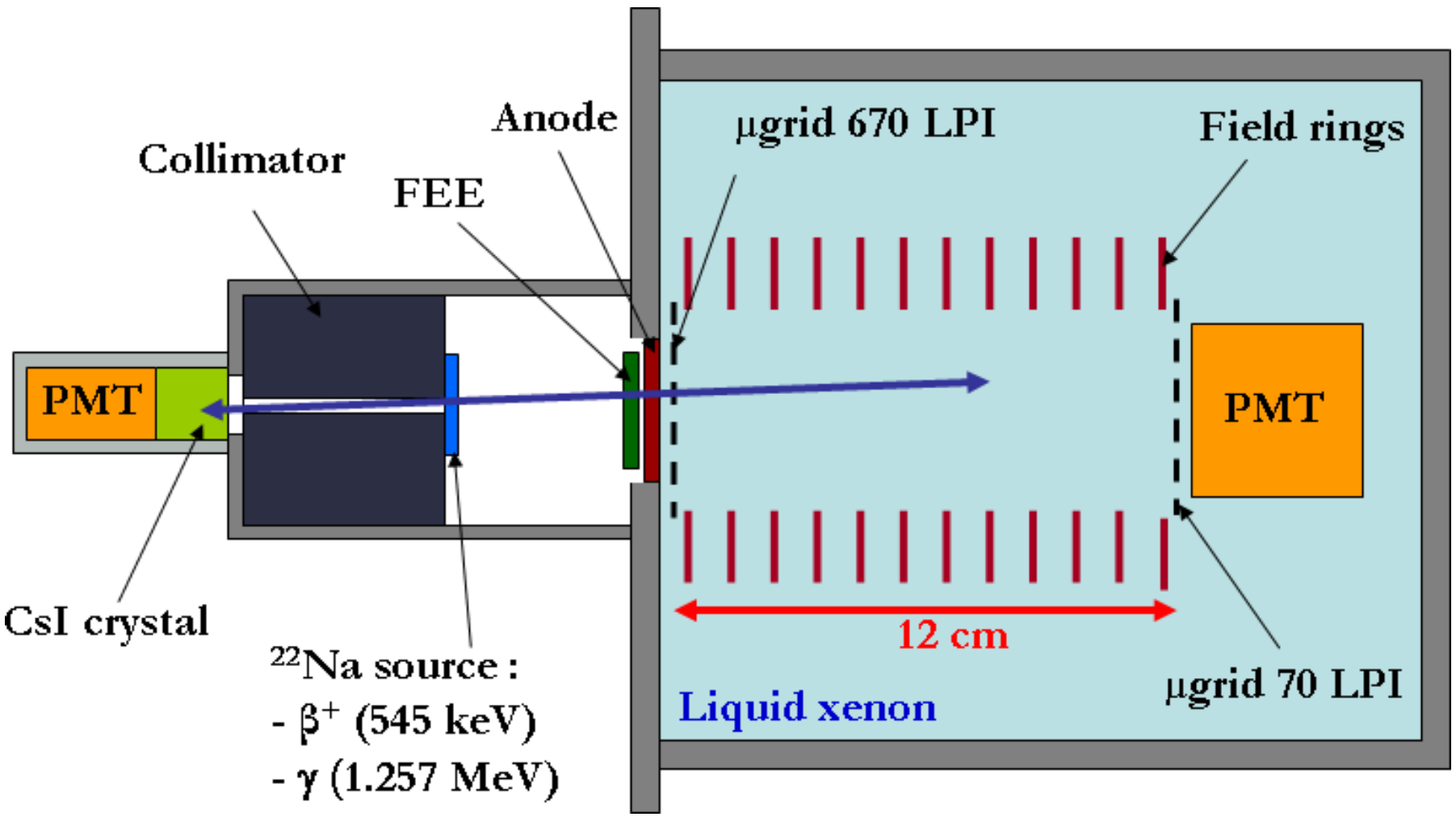

**Figure 3 : Diagram of the TPC characterization test-bench**

## 3. Results

Thanks to the waveform of PMT signal, we can determine the interaction time ($t_i$) since the decay time of LXe scintillation signal is small (2.2 ns); the x-y position, arrival time ($t_f$) and the amplitude of charge signals induced by drift electrons can also be determined by the waveform of 16-channel charge readout. Constant Fraction Discriminator (CFD) is applied for the determination of charge arrival time in order to improve drift time measurement precision. Figure 4 shows the drift time ($t = t_f - t_i$) distribution of single-hit events with 511 keV energy deposit, which is an exponential distribution with negative slope due to the cross-section of photoelectric effect. By fitting the distribution around t = 0 with an error function ($Erf(t/\sigma_t)$, where $\sigma_t$ is the time resolution due to the trigger dead-time, electronic noise and sampling frequency), $\sigma_t$ is estimated to be ~250 ns.

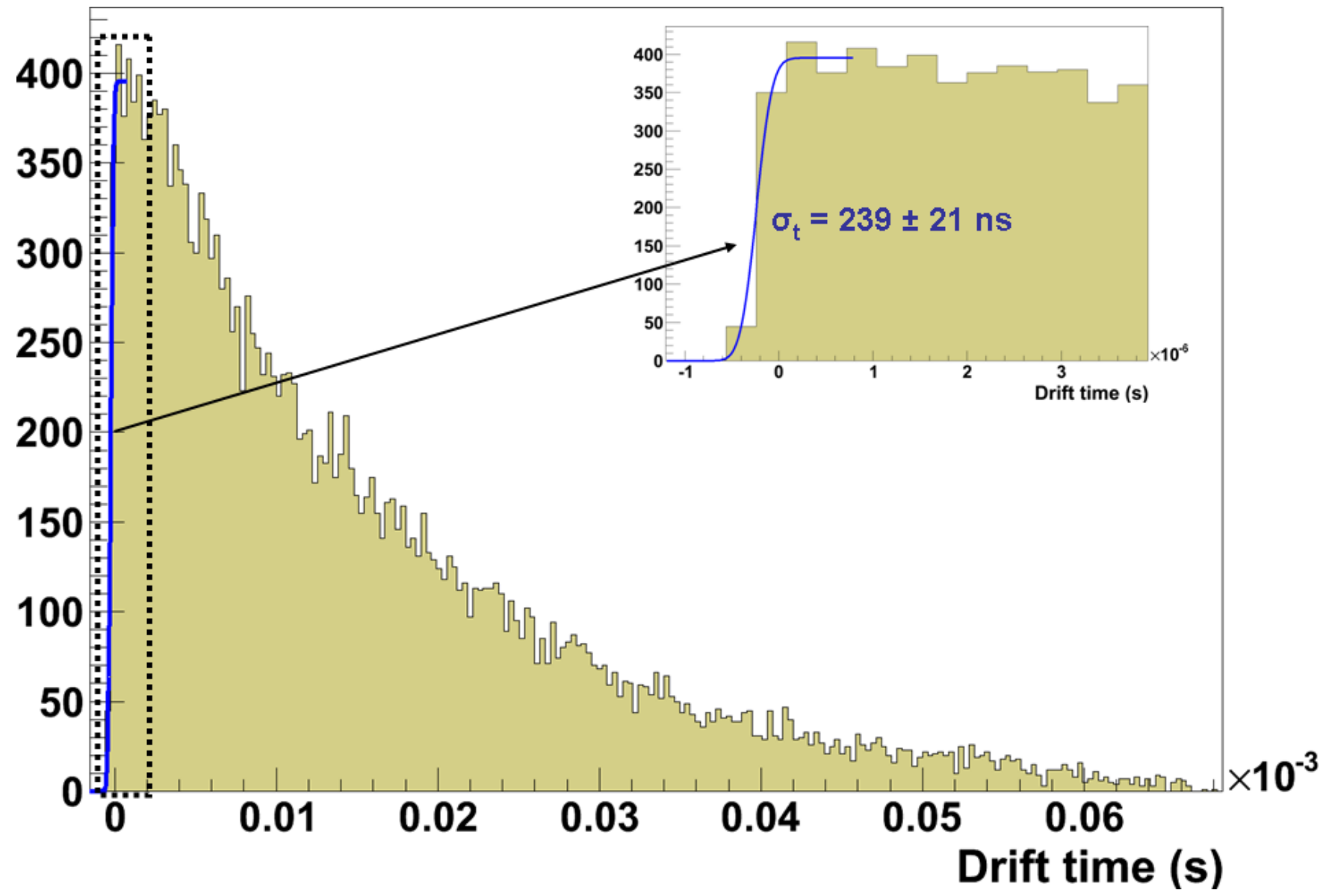

**Figure 4 : Distribution of the electron drift times of 511 keV gammas photoelectric interactions.**

The drift velocity of charge carriers as a function of the drift field is shown in Figure 5. For comparison, measurement made by Guschin et al. [7] (blue squares) and Yoshino et al. [8] (green stars) are presented. By multiplying the drift velocity and the time resolution we estimate a resolution along z smaller than 500 μm.

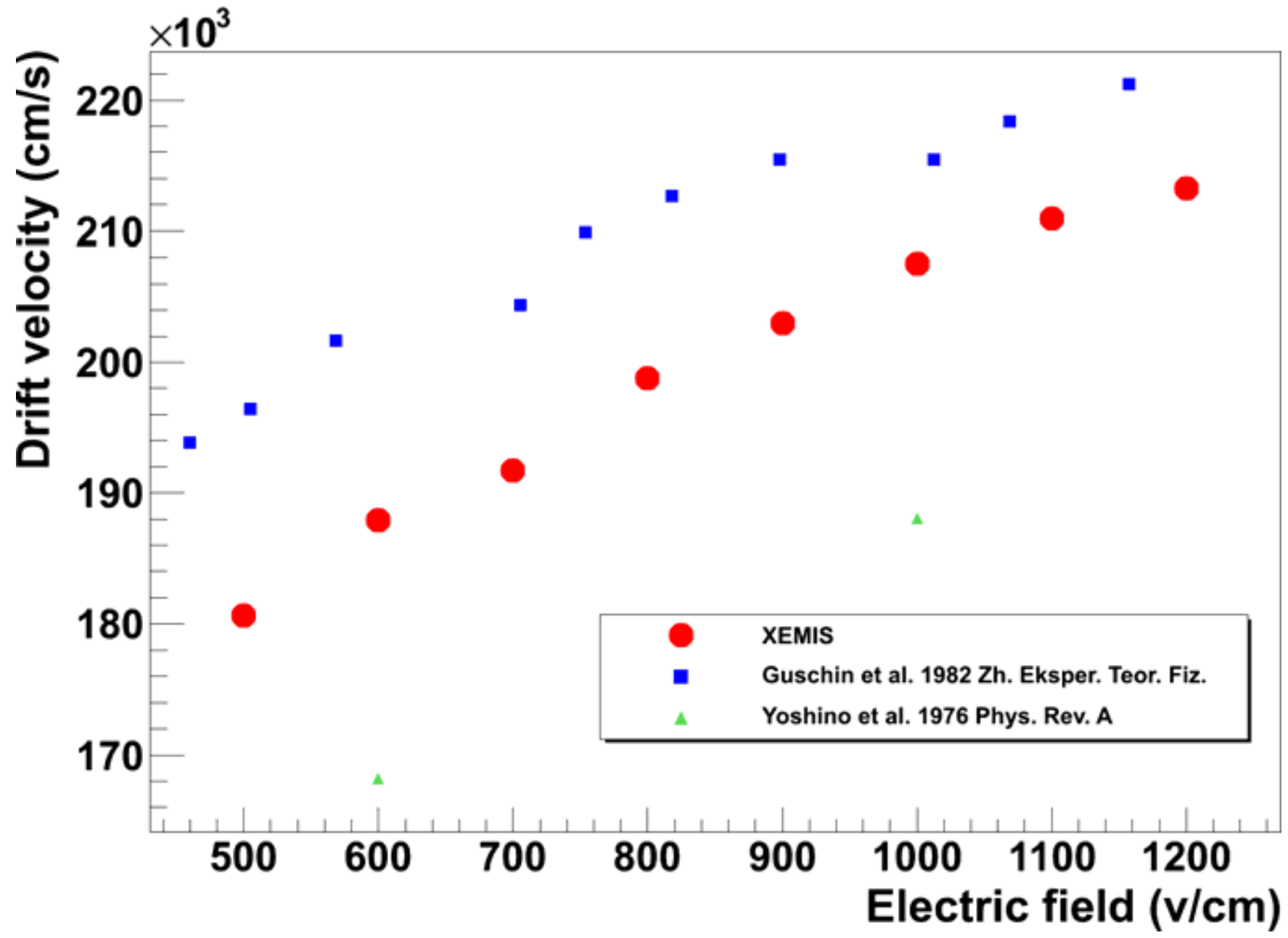

**Figure 5 : Electron drift velocity as a function of the electric drift field in liquid xenon. Errors bars of XEMIS measurements are within data points.**

Electronegative impurities worsen the charge yield by catching ionizing electrons along their drift. The probability that a drifting electron can encounter electronegative molecules before reaching the anode is related to the impurity level and drift length. Therefore the charge yield as a function of drift length S(z) can be expressed as (2),

$$S(z) = S_0 e^{\frac{-z}{\lambda}} \qquad (2)$$

where $S_0$ is the charge yield without attenuation, and $\lambda$ is the electron attenuation length, which is related to impurity level. To estimate the electron attenuation length, the amplitude of charge signal of 511 keV energy deposit as a function of z is measured at different time. $\lambda$ is estimated run-by-run by fitting the variation of amplitude with an exponential function, as shown in Figure 6. All of the measurements are made with an attenuation length higher than 15 cm, and the correction of the charge signals amplitude as a function of drift length is performed.

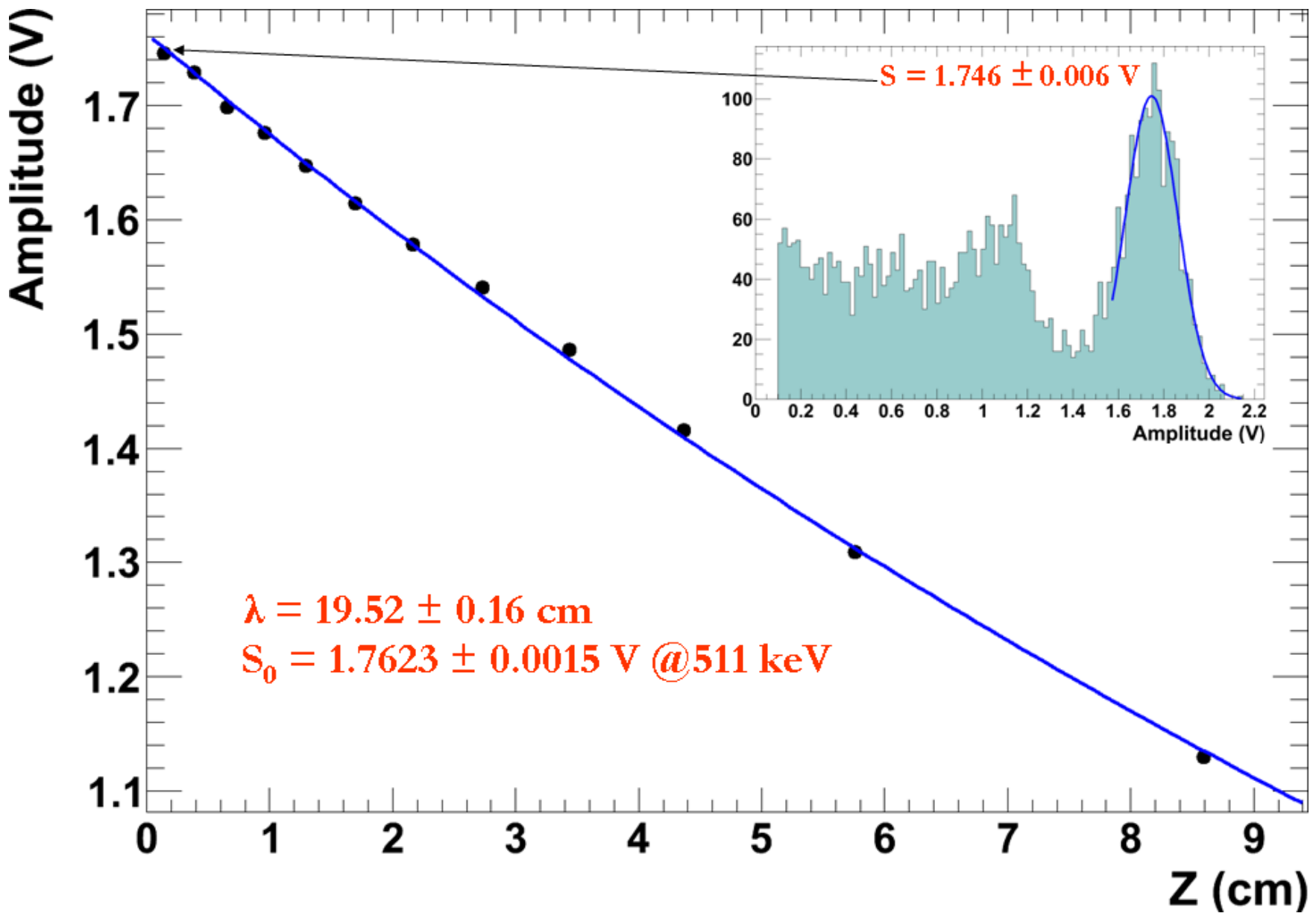

**Figure 6 : Signal amplitude of 511 keV photoelectric event as a function of the interaction depth in liquid xenon (z). On the top right corner, a typical energy spectrum is shown. The Gaussian fit of the photoelectric peak give the corresponding mean charge signal amplitude. Uncertainties are negligible.**

Figure 7 shows the corrected charge signal amplitude for 511 keV energy deposit as a function of the drift field. Since it was not possible to determine precisely the capacitance of the electronic at 171 K, we do not measure the absolute yield in electrons. We assumed as reference the yield predicted with Thomas model [9], 27 500 electrons at a drift field of 1 kV/cm. A fit of the curves plotted this model function modeling the collected charge is also shown (in red). This model is an expression of the collected charge (Q/Q0) (3):

$$\frac{Q}{Q_0} = a\frac{\ln(1+\xi_0)}{\xi_0} + (1-a)\frac{\ln(1+\xi_1)}{\xi_1} \qquad (3)$$

In the above equation, the parameter $\xi_0$ describes the recombination factor of the electron-ion pair in the minimum ionizing region of the particle track and $\xi_1$ is the recombination factor in the more heavily ionization density regions. The parameter *a* is a function of the threshold energy $E_1$ and $E_2$ which define the different behavior of delta electrons. Thanks to the model described in this reference, we are able to compare with other measurements, made at 570 keV by Aprile *et al.* [10] in green, 570 keV by Ichige *et al.* [11] in blue and 511 keV by Amaudruz *et al.* [12] in purple. By doing a linear extrapolation of our measurements to 570 keV, we find that our extrapolated points are between those made by Aprile *et al.* and Ichige *et al.* The discrepancy could be explained by the presence of systematic errors in the different measurements. With respect to Amaudruz *et al.* we observe a difference of 14 % at a drift field of 1 kV/cm.

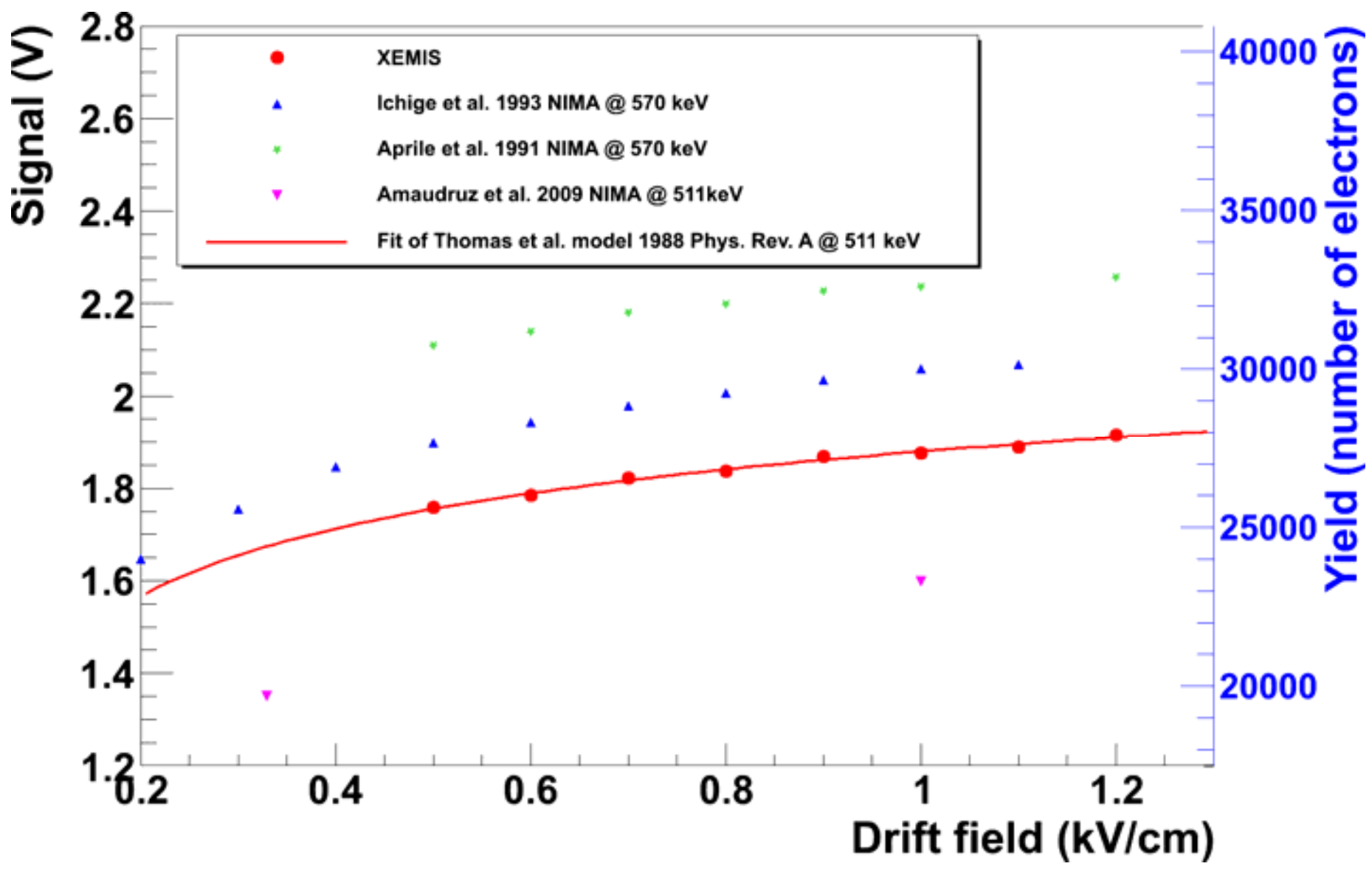

**Figure 7 : Charge yield of liquid xenon as a function of drift field for gamma-rays. Statistics uncertainties of XEMIS measurements are negligible.**

Energy resolution at 511keV as a function of drift field is presented in Figure 8. We see that energy resolution improve with electric drift field significantly. Observed resolutions are encouraging for further measurements with higher drift fields and different source energies.

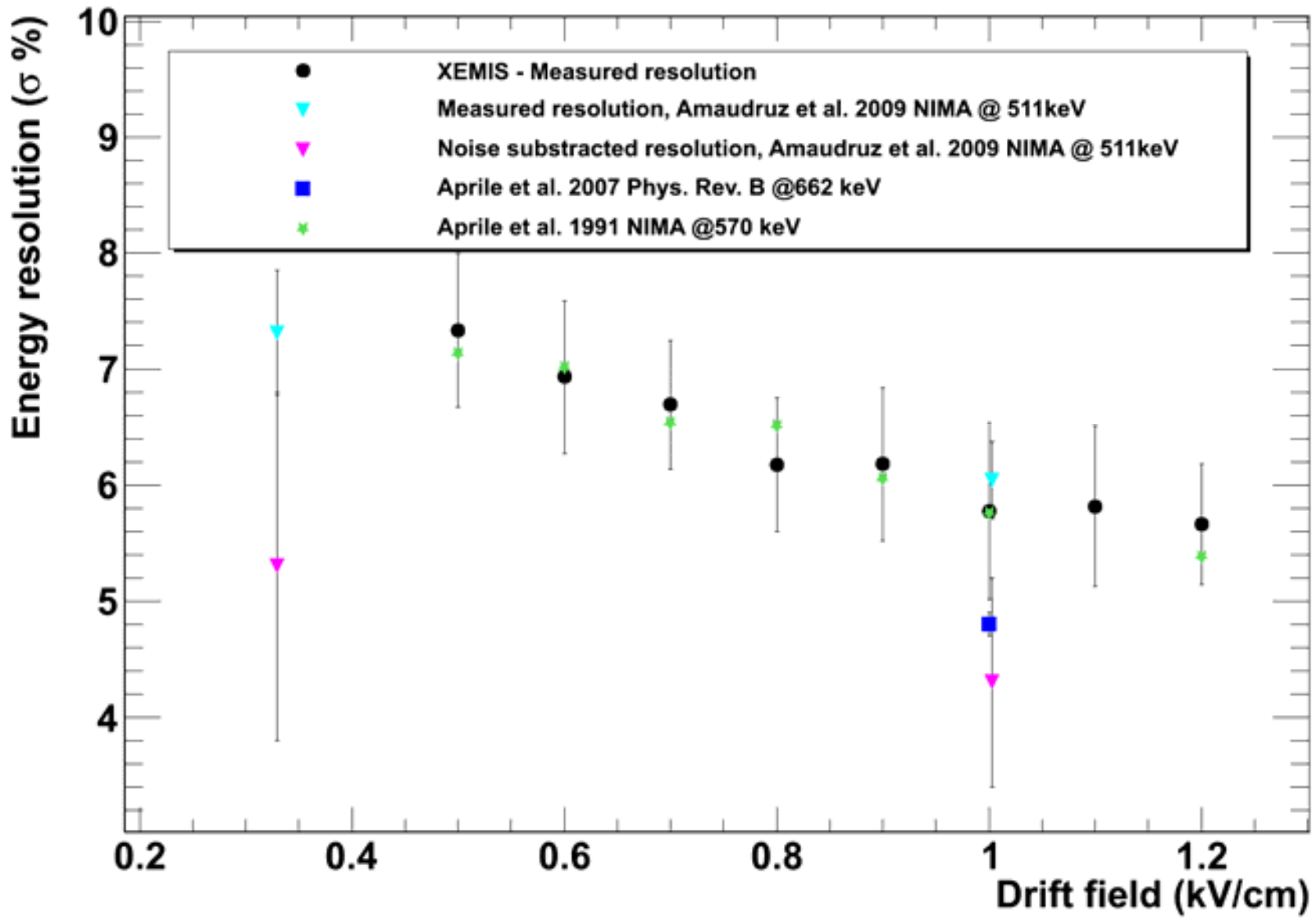

**Figure 8 : Energy resolution as a function of drift field**

## 4. Conclusion

A liquid xenon TPC containing a Micromegas ionization read-out and a low noise multi-channels front-end electronics has been successfully built and tested. We report on first results obtained with 511 keV γ-rays with the goal of investigating the new concept of the Compton “3γ imaging”. Very promising liquid xenon purity has been obtained and monitored in extremely satisfying conditions during long run operation. Very competitive time and energy resolutions are achieved. It is doubtless a strong step performed toward a new generation of liquid xenon Compton telescope and we hope toward a new medical imaging technique.

## Acknowledgments

Authors want to thanks Prof. Tomiyushi Haruyama for his help on the xenon cryogenics, Prof. Toshiaki Tauchi for very useful discussion on electronic read-out and Prof. Elena Aprile for her encyclopedic knowledge on xenon. This work is supported by the region of Pays de la Loire, France.